\documentclass[aps,prl,reprint,superscriptaddress]{revtex4-1}

\usepackage{graphicx}
\usepackage{amsmath}
\usepackage{amssymb}
\usepackage{placeins} %\FloatBarrier
\usepackage{siunitx}
\DeclareSIUnit{\abohr}{a_o}
\DeclareSIUnit{\erecoil}{E_r}
\DeclareSIUnit{\gauss}{G}

\newcommand{\ensavg}[1]{\left< #1 \right>} %ensemble average
\newcommand{\ket}[1]{\left\lvert #1 \right\rangle}%
\newcommand{\matrixel}[3]{\left< #1 \vphantom{#2#3} \right|
 #2 \left| #3 \vphantom{#1#2} \right>} % for Dirac matrix elements

\begin{document}

\title{Angle-resolved photoemission spectroscopy of a Fermi-Hubbard system}

\author{Peter T. Brown}
\affiliation{Department of Physics, Princeton University, Princeton, NJ 08544 USA}
\author{Elmer Guardado-Sanchez}
\affiliation{Department of Physics, Princeton University, Princeton, NJ 08544 USA}
\author{Benjamin M. Spar}
\affiliation{Department of Physics, Princeton University, Princeton, NJ 08544 USA}
\author{Edwin W. Huang}
\affiliation{Department of Physics, Stanford University, Stanford, CA 94305 USA}
\affiliation{Stanford Institute for Materials and Energy Sciences, SLAC National Accelerator Laboratory, Menlo Park, CA 94025 USA}
\author{Thomas P. Devereaux}
\affiliation{Stanford Institute for Materials and Energy Sciences, SLAC National Accelerator Laboratory, Menlo Park, CA 94025 USA}
\affiliation{Department of Materials Science and Engineering, Stanford University, Stanford, CA 94305 USA}
\affiliation{Geballe Laboratory for Advanced Materials, Stanford University, Stanford, CA 94305 USA}
\author{Waseem S. Bakr}
\email{wbakr@princeton.edu}
\affiliation{Department of Physics, Princeton University, Princeton, NJ 08544 USA}

\date{\today}

%%%%%%%%%%%%%%%%%%%%%%%%%%%%%%%%%%%
% Abstract
%%%%%%%%%%%%%%%%%%%%%%%%%%%%%%%%%%%

\begin{abstract}
Angle-resolved photoemission spectroscopy (ARPES) measures the single-particle excitations of a many-body quantum system with both energy and momentum resolution, providing detailed information about strongly interacting materials \cite{Damascelli2003}. ARPES is a direct probe of fermion pairing, and hence a natural technique to study the development of superconductivity in a variety of experimental systems ranging from high temperature superconductors to unitary Fermi gases.  In these systems a remnant gap-like feature persists in the normal state, which is referred to as a pseudogap \cite{Mueller2017}. A quantitative understanding of pseudogap regimes may elucidate details about the pairing mechanisms that lead to superconductivity, but developing this is difficult in real materials partly because the microscopic Hamiltonian is not known. Here we report on the development of ARPES to study strongly interacting fermions in an optical lattice using a quantum gas microscope. We benchmark the technique by measuring the occupied single-particle spectral function of an attractive Fermi-Hubbard system across the BCS-BEC crossover and comparing to quantum Monte Carlo calculations. We find evidence for a pseudogap in our system that opens well above the expected critical temperature for superfluidity. This technique may also be applied to the doped repulsive Hubbard model which is expected to exhibit a pseudogap at temperatures close to those achieved in recent experiments \cite{Mazurenko2017}.
\end{abstract} 

\maketitle

%%%%%%%%%%%%%%%%%%%%%%%%%%%%%%%%%%%
% Introduction
%%%%%%%%%%%%%%%%%%%%%%%%%%%%%%%%%%%

Photoemission spectroscopy measures the occupied single-particle spectral function \cite{Damascelli2003, Chen2009, Toermae2016}, which describes the allowed energies for a single-particle excitation with given momentum. ARPES has been used to study the presence of a Fermi surface, the lifetime of quasiparticles, superconducting gaps and their symmetries, and surface states in topological materials \cite{Vishik2010,Hasan2010}. It is illustrative to consider some generic features of the spectral function as we introduce interactions. For a non-interacting system with dispersion $\epsilon_k$, the single-particle excitations are eigenstates of the system implying they have a definite energy and infinite lifetime. The spectral function is a delta function, $A(k, \omega) = \delta(\omega - \epsilon_k + \mu)$ where $\mu$ is the chemical potential. Turning on weak interactions may create a Fermi liquid state, where the spectral function has a similar form but broadens along the energy axis reflecting the finite lifetime of the quasiparticles. In superconducting systems, more radical changes may occur including the development of a gap separating two disconnected spectral function branches. In weak coupling (BCS) superconductors, the gap vanishes at the critical temperature. However in many strongly-interacting superconducting systems including the high-temperature cuprate superconductors (HTSCs) and the unitary Fermi gas, a depression in the spectral weight at the chemical potential persists in the normal state \cite{Mueller2017, Chin2004, Schunck2007a, Gaebler2010, Nascimbene2011, Feld2011, Sommer2012, Murthy2017}. These so called pseudogap states might arise from the same pairing mechanism as the superconducting ground states, and developing a better understanding of their properties may shed light on pairing physics in the HTSCs.

Photoemission spectroscopy with cold atoms has been used to study the pseudogap regime in strongly interacting Fermi gases without a lattice. The idea was first proposed using Raman techniques \cite{Dao2007}, but was realized using momentum-resolved radiofrequency (rf) spectroscopy \cite{Stewart2008}. Previous experiments have explored the BCS-BEC crossover in 3D \cite{Stewart2008, Gaebler2010, Sagi2015} and 2D \cite{Feld2011, Frohlich2012} continuum systems. These experiments inferred the presence of a pseudogap from spectral features that exhibit ``back-bending'' near the Fermi surface at temperatures above $T_c$, in analogy to the back-bending observed in the BCS dispersion. This interpretation has been criticized because back-bending can also arise due to other effects, including universal short range physics expected in continuum systems \cite{Schneider2010}. Exact theory techniques were not available for these systems, therefore experiments compared with various $T$-matrix approximation schemes \cite{Loktev2001, Perali2002, Perali2011} which become exact for weak interactions. 

%%%%%%%%%%%%%%%%%%%%%%%%%%%%%%%%%%%
% Figure 1
%%%%%%%%%%%%%%%%%%%%%%%%%%%%%%%%%%%

\begin{figure*}[ht]
\centering
\includegraphics[width=0.75\textwidth]{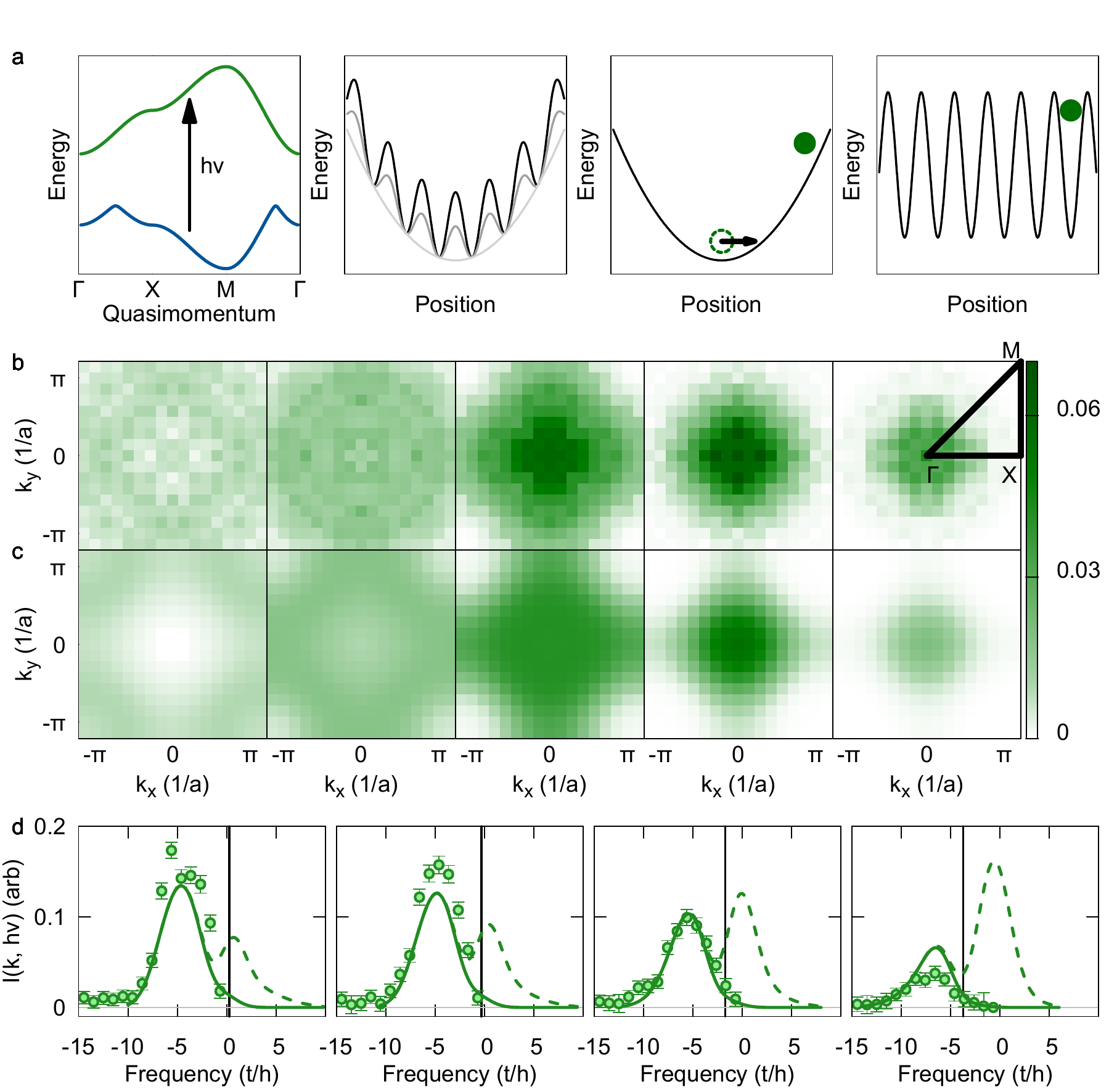}
\caption{{\bfseries ARPES technique and raw data.} {\bfseries a}, A radiofrequency photon of energy $h \nu$ is incident on an interacting system  with a dispersion (blue) and excites state $\ket{\uparrow}$ atoms to state $\ket{f}$ with a non-interacting final dispersion shown in green (far left). The dispersions are plotted along the high-symmetry lines of the Brillouin zone, with $\Gamma = (0, 0)$, $X = (\pi, 0)$, and $M = (\pi, \pi)$. We perform band mapping to map quasimomentum to momentum (second from left) by ramping down the lattice depth adiabatically with respect to the band gap. The atoms expand for a quarter period in a harmonic trap (third from left). Atoms (green ring) with initial momentum $\hbar k$ expand to position $r = k/l^2$ (solid green circle). Finally, we freeze the position of the atoms by ramping up an optical lattice to $\sim \SI{60}{\erecoil}$ (right-most). {\bfseries b}, Atomic density after the quarter period expansion for a range of frequencies $\nu$ at $U/t=-7.5(1)$ and $T/t = 0.55(3)$. From left to right, $h(\nu - \nu_o)/t = -10.5$, $-8.6$, $-6.6$, and $-3.6$ and $-1.7$. Average of ~40 pictures, binned, and averaged using the symmetry of the square. Field of view of $\sim \SI{40}{\micro \meter} \times \SI{40}{\micro \meter}$. {\bfseries c}, DQMC results for $A(k, \epsilon_k - h\nu) F(\epsilon_k - h\nu)$ at $U/t = -7.5$ and $T/t = 0.55$ for the same values of $\nu$ shown in b. {\bfseries d}, Experimental (points) and DQMC (solid green lines) results for the occupied spectral function, DQMC results for the full spectral function (dashed lines), and chemical potential (black lines) at several quasimomenta in the Brillouin zone. From left to right, $(k_x, k_y) = (0, 0)$, $(\pi/4, 0)$, $(\pi/2, 0)$, and $(\pi, 0)$. Error bars SEM.}
\label{fig:raw_data}
\end{figure*}

%%%%%%%%%%%%%%%%%%%%%%%%%%%%%%%%%%%
% Figure 2
%%%%%%%%%%%%%%%%%%%%%%%%%%%%%%%%%%%

\begin{figure*}[ht]
\centering
\includegraphics[width=0.75\textwidth]{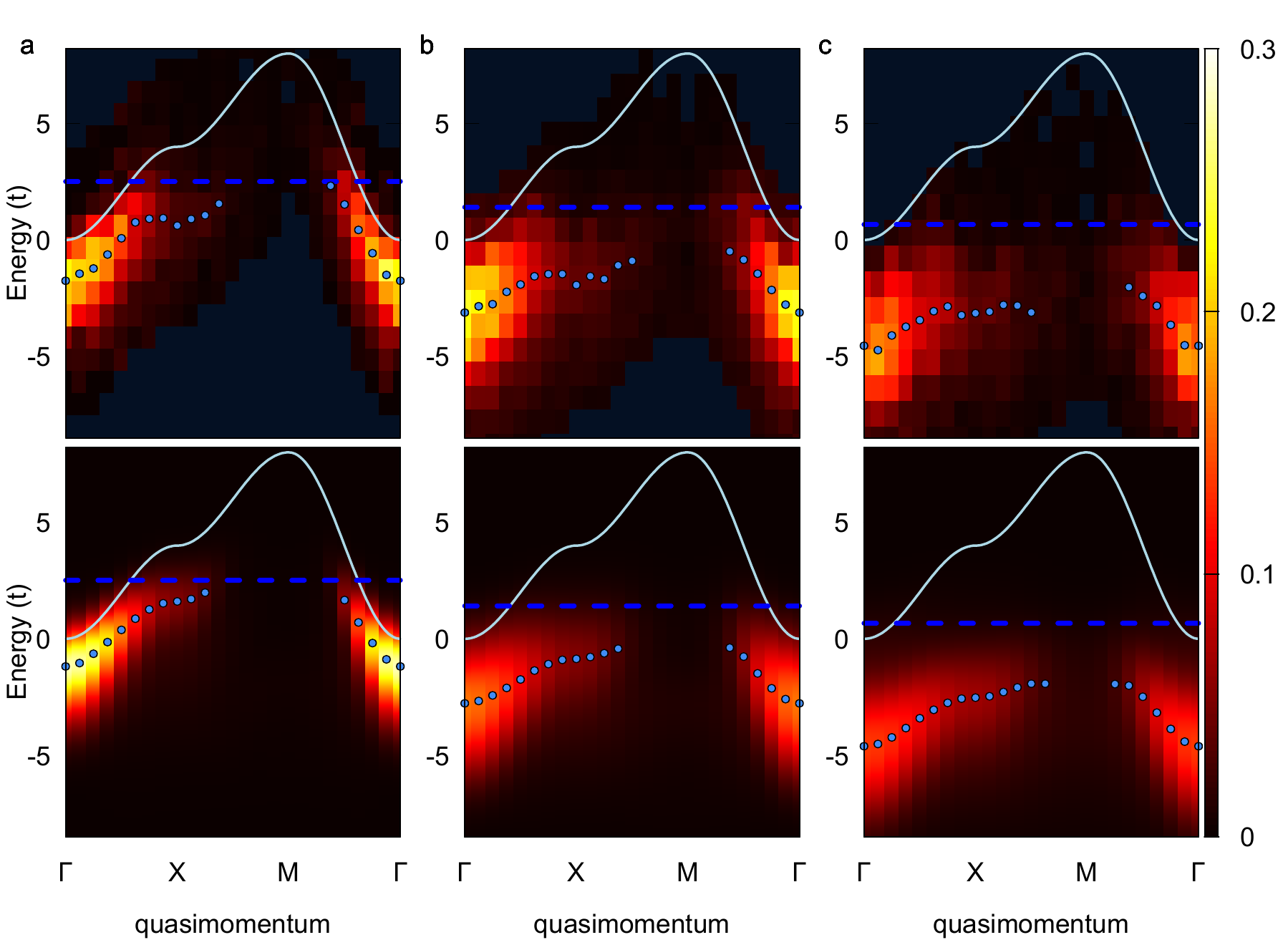}
\caption{ {\bfseries Occupied spectral function versus interaction.}  {\bfseries a}, Experimental occupied spectral function (top) along directions of high symmetry in the Brillouin zone for $U/t=-3.7(1)$ and $T/t = 0.48(2)$ with the center position of Lorentzian fits to each EDC (light blue circles), the chemical potential (blue line) and the non-interacting dispersion relation (white line).  DQMC results (bottom) for $U/t = -4$ and $T/t = 0.5$. {\bfseries b}, Experiment (top) for $U/t = -6.0(1)$ and $T/t = 0.50(2)$ and DQMC (bottom) for $U/t = -6$ and $T/t = 0.5$. {\bfseries c} experiment (top) for $U/t = -7.5(1)$ and $T/t = 0.55(3)$ and DQMC (bottom) for $U/t = -7.5$ and $T/t = 0.55$.} 
\label{fig:vs_interaction}
\end{figure*}

%%%%%%%%%%%%%%%%%%%%%%%%%%%%%%%%%%%
% Figure 3
%%%%%%%%%%%%%%%%%%%%%%%%%%%%%%%%%%%

\begin{figure*}[ht]
\centering
\includegraphics[width=\textwidth]{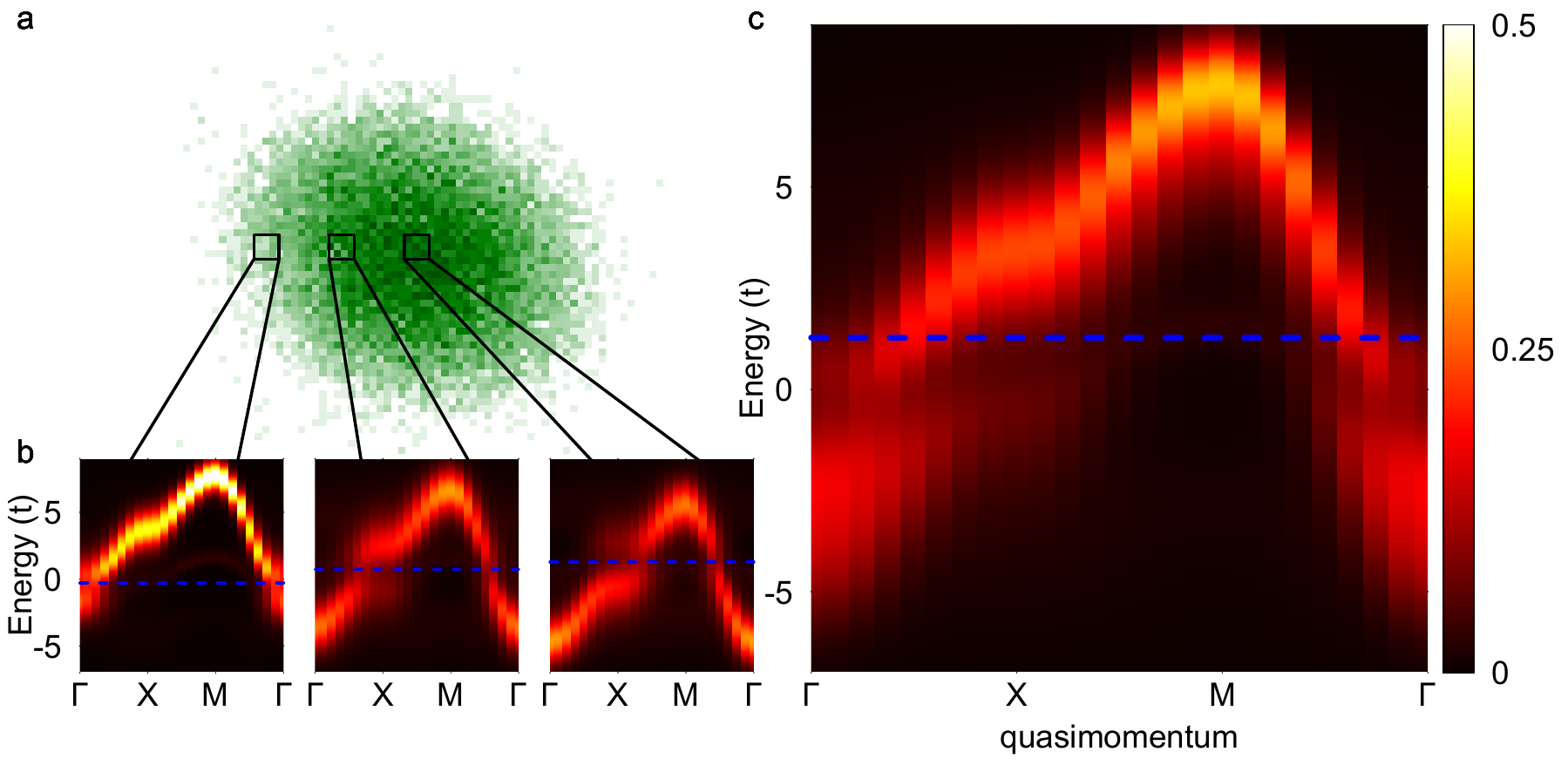}
\caption{ {\bfseries Trap-averaged spectral function.}  {\bfseries a}, The total density, $\ensavg{n}$, of the trapped atoms. Field of view $\sim \SI{50}{\micro \meter} \times \SI{50}{\micro \meter}$. Black squares indicate regions of the trap with densities $n = 0.2$, $0.8$, and $1.2$ from left to right. {\bfseries b}, Homogeneous system spectral functions calculated using DQMC at $U/t = -6$, $T/t = 0.5$ and $n = 0.2$ (left), $0.8$ (center), and $1.2$ (right). The chemical potential at each density is denoted by a horizontal blue line. {\bfseries c}, The trap average spectral function constructed by weighting the homogeneous system spectral functions by the area of the trapped cloud at each density.} 
\label{fig:trp_avg}
\end{figure*}

%%%%%%%%%%%%%%%%%%%%%%%%%%%%%%%%%%%
% Figure 4
%%%%%%%%%%%%%%%%%%%%%%%%%%%%%%%%%%%

\begin{figure*}[ht]
\centering
\includegraphics[width=\textwidth]{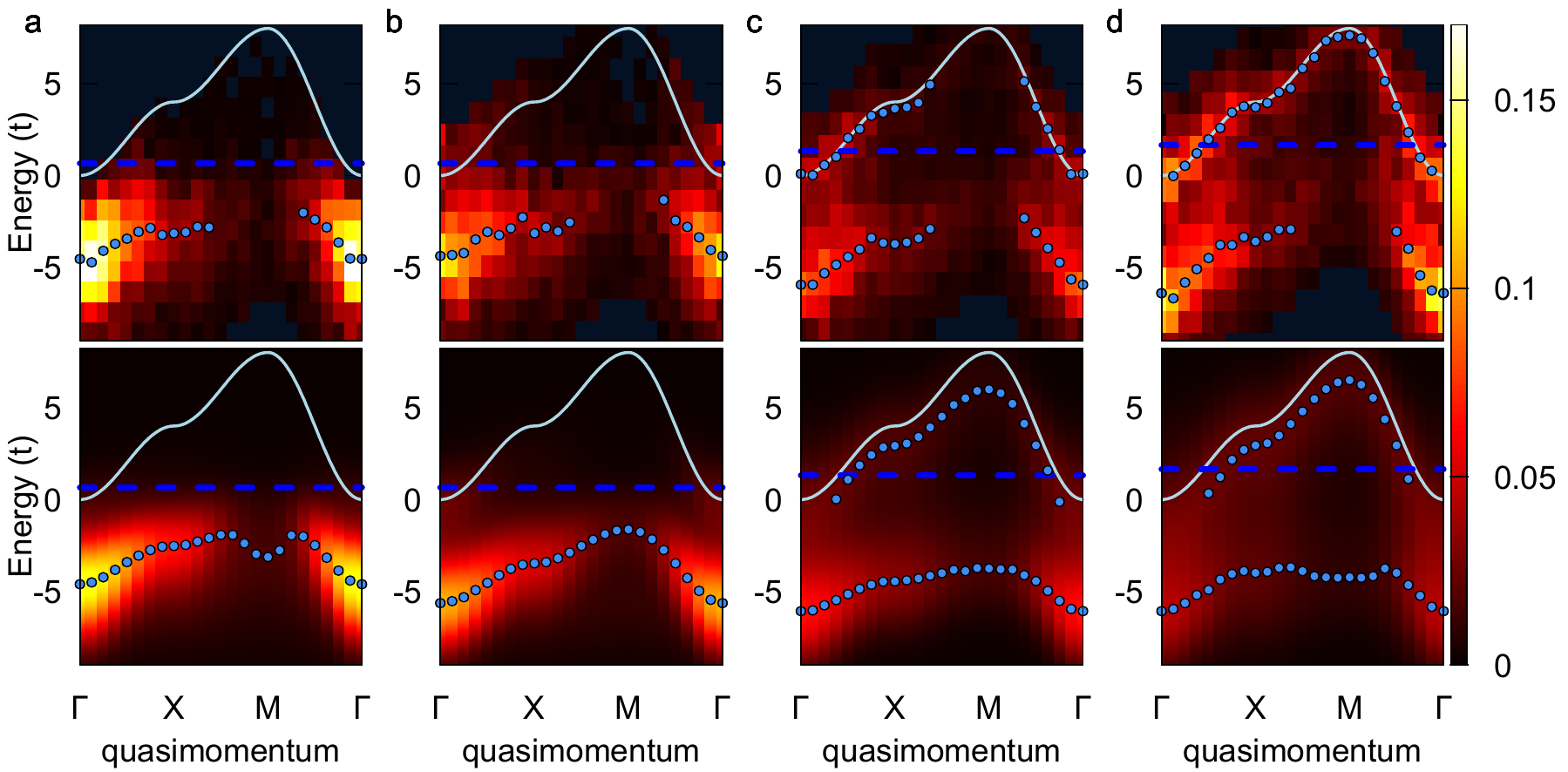}
\caption{ {\bfseries Occupied spectral function versus temperature at strong coupling.}  {\bfseries a} Experimental occupied spectral function along directions of high symmetry in the Brillouin zone (top) for $U/t = -7.5(1)$ and $T/t = 0.55(3)$ and DQMC (bottom) for $U/t = -7.5$ and $T/t = 0.55$. Also shown are the location of peaks in the occupied spectral function (light blue circle), the chemical potential (blue line), and the non-interacting dispersion (white line). {\bfseries b}, Experiment (top) for $U/t = -7.5$ and $T/t = 0.93(5)$ and DQMC (bottom) for $U/t = -8$ and $T/t = 1.04$. {\bfseries c} Experiment (top) for $U/t = -7.5$ and $T/t = 2.5(1)$ and DQMC (bottom) for $U/t = -8$ and $T/t = 2.83$. {\bfseries d}, Experiment (top) for $U/t = -7.5$ and $T/t = 4.5(1)$ and DQMC (bottom) for $U/t = -8$ and $T/t = 5$.} 
\label{fig:vs_temp}
\end{figure*}

Here we extend photoemission spectroscopy to strongly-correlated lattice systems. We take advantage of quantum gas microscopy techniques to obtain high signal-to-noise data using a single layer 2D system, avoiding the complication of integration over a third axis of momentum space. Combining photoemission spectroscopy with quantum gas microscopy poses a technical challenge because photoemission spectroscopy relies on a large field of view for time-of-flight measurements to reach momentum space, whereas quantum gas microscopy is limited to a small field of view over which the atoms can be pinned during imaging. We solve this problem by developing a new measurement protocol which consists of four parts, illustrated in fig.~\ref{fig:raw_data}a. First, we excite atoms from the interacting system to a non-interacting final state using a radiofrequency pulse. Next, we perform band mapping to adiabatically connect quasimomentum states in the lattice to real momentum states. Then, we allow the atoms to expand in a harmonic trap for a quarter period which maps the atomic momentum distribution to the real space distribution \cite{Murthy2014}. Finally, we turn on a deep optical lattice to freeze the atoms in place in preparation for site resolved imaging. The momentum detection portion of this protocol has been discussed in a recent proposal for measuring the spectral function in 1D systems using lattice modulation techniques \cite{Bohrdt2018}. 

We benchmark the ARPES technique by studying the pseudogap regime in the attractive Hubbard model. There has been comparatively little experimental work on this system \cite{Strohmaier2007,Hackermuller2010,Schneider2012,Mitra2017}, but it has a number of advantages as a testing ground for our new protocol. Unlike the repulsive model, the attractive model does not have a sign problem, meaning it is possible to make high-precision theory calculations using DQMC at low temperatures. Furthermore, the 2D attractive model exhibits phenomenology similar to other strongly interacting superconducting systems. In particular, it exhibits a BCS-BEC crossover with increasing on-site interaction strength and a Berezinskii-Kosterlitz-Thouless (BKT) transition to an $s$-wave superconducting state at low temperatures \cite{Paiva2004}. At weak interactions the ground state is BCS-like, and pairing is destroyed at the critical temperature for superconductivity, $T_c$. As the interaction strength is increased, pairing persists above the critical temperature and is destroyed at some higher temperature $T^* > T_c$. Intermediate temperatures, $T_c < T < T^*$, exhibit a pseudogap which appears as a depression in either the density of states or spectral function at the Fermi energy \cite{Singer1996, Singer1998, Singer1999}. As the interaction strength increases, the pairing changes character from momentum-space to real-space and the pseudogap temperature increases monotonically. In the BEC limit, the pseudogap represents a regime of real-space bosonic pairs which develop phase coherence below $T_c$.
 
%%%%%%%%%%%%%%%%%%%%%%%%%%%%%%%%%%%
% Experiment description
%%%%%%%%%%%%%%%%%%%%%%%%%%%%%%%%%%%

In this experiment, we study an attractive Hubbard system with interactions ranging from intermediate coupling, $U/t = -4$, to strong coupling, $U/t = -8$. We observe the emergence of a pseudogap as we increase the interaction at fixed temperature, $T/t \sim 0.5$. This is well above the theoretically calculated maximum $T_c$ for this model $T/t \sim 0.15$ \cite{Paiva2004}. At strong coupling and hotter temperatures, we observe the presence of two branches in the occupied spectral function due to the coexistence of single and paired atoms.

To begin, we prepare an equal mixture of hyperfine states $\ket{\uparrow} = \ket{3}$ and $\ket{\downarrow} = \ket{1}$ in a 2D trap at a field of $\sim\SI{120}{\milli \gauss}$ where the scattering length $a_{13} \sim \SI{-800}{\abohr}$ \cite{O'Hara2000}, where \si{\abohr} is the Bohr radius and $\ket{i}$ denotes the $i$th lowest hyperfine state of lithium. We adiabatically load a 2D optical lattice with lattice spacing $a = \SI{752}{\nano \meter}$ to depths between $5$ and \SI{10}{\erecoil}, where $\si{\erecoil}/h = \SI{14.7}{\kilo \hertz}$ is the lattice recoil energy. We then apply a \SI{1}{\milli \second} radiofrequency (rf) pulse to transfer $\sim 15\%$ of the $\ket{\uparrow}$ atoms to a third state $\ket{f} = \ket{2}$ (fig.~\ref{fig:raw_data}a, left panel). Final state interactions are negligible because the scattering lengths $a_{12} \sim \SI{0}{\abohr}$ and $a_{23} \sim \SI{25}{\abohr}$ are small. We determine the initial quasimomentum distribution of the probed atoms by performing band mapping and subsequently expanding for a quarter period in a harmonic trap. To achieve this, we first turn on a harmonic trapping potential with trapping frequency $\omega = (2\pi) \SI{360}{\hertz}$ in a few microseconds. Then, we ramp the lattice potential to zero linearly in \SI{200}{\micro \second} before expanding for $T/4 = \SI{700}{\micro \second}$. The real space density distribution after the quarter period expansion reflects the initial quasimomentum distribution of the $\ket{f}$ atoms according to $\tilde{n}_f(k) = n_f(r) l^4$, where $l = \sqrt{\hbar/m\omega}$ is the harmonic oscillator length and $k = r / l^2$.

%%%%%%%%%%%%%%%%%%%%%%%%%%%%%%%%%%%
% Raw data description
%%%%%%%%%%%%%%%%%%%%%%%%%%%%%%%%%%%

We obtain the energy dependence of the spectral function by measuring the density profile at a range of rf frequencies. For each frequency, $\nu$, we obtain a two-dimensional image, $n_f(r, \nu)$ as shown in fig.~\ref{fig:raw_data}b. Each pixel in the image is associated with a different quasimomentum. To improve our signal-to-noise, we spatially average using the symmetry of the square. The final density distribution of state $\ket{f}$ atoms is proportional to the occupied single-particle spectral function, $n_f(r, \nu) \propto A(\mathbf{k}, \epsilon_\mathbf{k} - h \nu - \mu) F(\epsilon_\mathbf{k} - h \nu - \mu)$ \cite{supplement}, where $F$ is the Fermi function. The rf photon carries negligible momentum, therefore the initial momentum is the same as the final momentum. The energy of the final state is given by the tight-binding dispersion $\epsilon_\mathbf{k}/t = 4 - 2 \cos(k_x a) - 2 \cos(k_y a)$, where $t$ is the tunneling energy, and the rf photon energy $h \nu$ leading to an initial energy $\epsilon_\mathbf{k} - h \nu$. This transition is illustrated schematically in momentum space (fig.\ref{fig:raw_data}a, left panel).

The density profiles change shape as the rf frequency is scanned, a characteristic of an interacting system. At large red detuning relative to the non-interacting transition, the transferred density is more heavily weighted towards the edges of the the Brillouin zone (fig.~\ref{fig:raw_data}b, leftmost). As the detuning is decreased, the density becomes peaked at the center and the transfer increases (center). Decreasing the detuning further leads to less transfer but does not significantly change the shape (rightmost). To further quantify the energy dependence of the spectral function we focus on individual quasimomenta and plot the corresponding energy distribution curves (EDCs) in fig.~\ref{fig:raw_data}d. We compare our experimental results with determinantal quantum Monte Carlo (DQMC) calculations at $U/t = -7.5$ and $T/t = 0.55$, which are shown in fig.~\ref{fig:raw_data}c and d and find good agreement. The full DQMC spectral function (fig.~\ref{fig:raw_data}d) exhibits a depression near the central chemical potential, indicating a pseudogap at this temperature and interaction.

%%%%%%%%%%%%%%%%%%%%%%%%%%%%%%%%%%%
% Results
%%%%%%%%%%%%%%%%%%%%%%%%%%%%%%%%%%%
Having demonstrated the viability of this photoemission spectroscopy technique, we focus on elucidating the properties of the attractive Hubbard system by exploring the evolution of the spectral function with interaction strength and temperature. First, we consider the low temperature regime $T/t \sim 0.5$ for a range of interactions $U/t = -4$, $-6$, and $-7.5$. For each interaction strength we determine the temperature, interaction, and global chemical potential by fitting in-situ densities and correlators to DQMC results \cite{supplement}. We determine the occupied spectral function from the density according to $A_\text{occ}(k, \omega) = n_f(k, \epsilon_k - \omega - \mu)$ and obtain the profiles shown in fig.~\ref{fig:vs_interaction}. For the weakest interaction we consider, $U/t=-4$, the spectral weight is shifted to negative energy relative to the non-interacting dispersion and there is significant spectral weight near the chemical potential. As we increase the interaction, the spectral weight shifts further from the non-interacting dispersion and there is less spectral weight near the chemical potential.

We quantify the distribution of spectral weight by fitting each EDC to a Lorentzian profile and extracting the center position and half width half maximum (HWHM) values. We emphasize that the resulting graph cannot be interpreted as a dispersion for two reasons. First, there are trap averaging effects and second even in a homogeneous system the Fermi function shifts the position of the peak in $A_\text{occ}(k, \omega)$ relative to $A(k, \omega)$, causing significant deviation from the true dispersion near and above the Fermi surface \cite{supplement}. For $U/t =-4$, the center of the spectral weight reaches the chemical potential at a point along the $X-M$ line in momentum space (fig.~\ref{fig:vs_interaction}a). For stronger interactions, $U/t = -6, -7.5$ the center of spectral weight no longer approaches the chemical potential. This pulling back of the spectral weight at the chemical potential is a signature of pseudogap behavior. The spectral width increases weakly with the interaction. At $U/t = -4$, the HWHM is $\sim 1.2 t$ with little dependence on the quasimomentum. As we increase the interaction the width increases to $\sim 2 t$ and $\sim 2.5 t$ at $U/t = -6$ and $U/t = -7.5$ respectively. The measured width is larger than the Fourier broadening which is $\SI{1}{\kilo \hertz}$, corresponding to $\sim 0.7 t$, $\sim 0.8 t$, and $\sim 1 t$ for these interactions. The width is not intrinsic but due to the spread of chemical potentials across the harmonic trap \cite{supplement}.

We compare our measurements with DQMC calculations which account for both the presence of the trap and Fourier broadening of the rf pulse. These results, shown in the lower panels of fig.~\ref{fig:vs_interaction}, display similar trends to the experiment. We determine the proportionality constant between the experimental data and the theoretical spectral function using a least squares fit. After scaling the experimental data, the theory agrees well at low quasimomentum \cite{supplement}. At larger quasimomentum near the $X$ point, the experimental spectral function is systematically smaller than the theory. This may be due to imperfections in the band mapping process which are more significant near the edge of the band. A pseudogap appears as a depression in the spectral function at the chemical potential in the theory results for $U/t = -6$ and $-8$ \cite{supplement}. The pseudogap is not present at $U/t = -4$, consistent with the calculations of \cite{Singer1998}. We determine the peak locations using a peak finding algorithm, which identifies multiple peaks more efficiently in the high quality numerical data.

The spectral function is broadened by the presence of the harmonic trapping potential which causes the system to sample a range of chemical potentials. We account for this effect in our theory results using the local density approximation. The trap spectral function is given in terms of the homogeneous system spectral functions by 
\begin{equation}
A_\text{trap}(k, \omega) = \frac{1}{N} \sum_i A(k, \omega + \mu_i - \mu, \mu_i) \label{eq:trp_avg_spec_fn},
\end{equation}
where $N$ is the number of sites in our trap, $\mu$ is the global chemical potential, and $\mu_i$ is the local chemical potential at site $i$. The trap causes broadening of spectral features and tends to obscure signatures of gaps or pseudogaps. We illustrate this effect for an experimental density profile in fig.~\ref{fig:trp_avg}. The theoretical homogeneous system spectral functions shown in fig.~\ref{fig:trp_avg}b exhibit a pseudogap for densities $n > 0.2$. When these functions are combined using eq.~\ref{eq:trp_avg_spec_fn}, the pseudogap feature is broadened over the spread of chemical potentials where the uniform system exhibits a pseudogap (fig.~\ref{fig:trp_avg}c). This effect is also visible in fig.~\ref{fig:raw_data}c, where the pseudogap feature appears shifted away from the central chemical potential. Before comparing these results with experiment, we account for Fourier broadening of the rf pulse by convolving the trap averaged spectral function with a Gaussian with standard deviation $h \times \SI{1}{\kilo \hertz}$.

We also consider the evolution of the spectral function with increasing temperature at the strongest interaction, $U/t = -7.5$. At the lowest temperature we find little weight at the Fermi level, fig.~\ref{fig:vs_temp}a. As the temperature is increased, a second peak emerges which approximately follows the non-interacting dispersion, fig.~\ref{fig:vs_temp}b,c,d. At the same time, the lower peak shifts to more negative energy and the spectral function is gapped at all momenta. At the hottest temperatures (fig.~\ref{fig:vs_temp}c,d), the Fermi weighting approaches a constant value and occupied spectral function is proportional to the full spectral function, $A_\text{occ}(k,\omega) \sim A(k,\omega)/2$.  The upper feature is due to unpaired atoms and the lower feature due to pairs.

We fit the EDCs to either one Lorentzian for the coldest temperatures or two Lorentzians for the hotter temperatures. We find that temperature has a weak effect on the peak widths. The HWHMs of the upper and lower peak are $\sim 1.5t$ and $\sim 2.5 t$ respectively. The theory results show that the dispersion obtained from the occupied spectral function nearly coincides with that obtained for the true spectral function $T/t \sim 0.5 - 3$. They also show that a pseudogap is present at $U/t = -8$ up to $T/t = 5$ \cite{supplement}. The large value of $T^*$ is characteristic of being in the strong coupling regime.

%%%%%%%%%%%%%%%%%%%%%%%%%%%%%%%%%%%
% Conclusion
%%%%%%%%%%%%%%%%%%%%%%%%%%%%%%%%%%%

In conclusion, we have developed a technique to measure the single-particle spectral function in Hubbard systems using a quantum gas microscope. In combination with recent transport experiments \cite{Brown2019, Nichols2019}, this extends the toolkit for measuring dynamical properties of these systems, which is so far rather limited compared with the techniques available to study HTSCs and other real materials. Our experiment paves the way for future studies of the single-particle spectral function in the doped repulsive Hubbard model, a quantity which is difficult to calculate using exact theory techniques because of the fermionic sign problem. Such an experiment might elucidate properties of the Hubbard system at high temperature where dynamical mean field theory studies have suggested quasiparticles survive into the linear resistivity regime \cite{Xu2013, Deng2013} or test recent parton theories which may describe the doped antiferromagnetic regime \cite{Grusdt2018}. At lower temperatures currently beyond the reach of exact theory methods, this technique could be used to search for the symmetry of the pseudogap and superconducting states.

\section{Acknowledgments}

\begin{acknowledgments}
 This work was supported by the NSF (grant no. DMR-1607277), the David and Lucile Packard Foundation (grant no. 2016-65128), and the AFOSR Young Investigator Research Program (grant no. FA9550-16-1-0269).  TPD and EWH acknowledge support from the U.S. Department of Energy, Office of Science, Office of Basic Energy Sciences, Division of Materials Sciences and Engineering, under Contract No. DE-AC02-76SF00515. Computational work was performed on the Sherlock cluster at Stanford University.
\end{acknowledgments}

%%%%%%%%%%%%%%%%%%%%%%%%%%%%%%%%%%%
% Bibliography
%%%%%%%%%%%%%%%%%%%%%%%%%%%%%%%%%%%

%

%%%%%%%%%%%%%%%%%%%%%%%%%%%%%%%%%%%
% Supplement
%%%%%%%%%%%%%%%%%%%%%%%%%%%%%%%%%%%

\pagebreak
\clearpage
\setcounter{equation}{0}
\setcounter{figure}{0}

\renewcommand{\theparagraph}{\bf}
\renewcommand{\thefigure}{S\arabic{figure}}
\renewcommand{\theequation}{S\arabic{equation}}

\onecolumngrid
\flushbottom

\section{Supplementary Material}

\section{The single-particle spectral function}

The single-particle spectral function comes from the single-particle retarded Green's function, which is defined by
\begin{eqnarray}
G^R(k, t) &=& -i \theta(t) \ensavg{\left[c_k(t), \ c^\dag_k(0) \right]_+}.
\end{eqnarray}

This function gives information about the single-particle excitations of a many body system. The first term in the anticommutator, $\ensavg{c_k(t) c^\dag_k(0)}$, describes what happens if we create a particle excitation with momentum $k$, watch this excitation propagate until time $t > 0$, and then measure what portion of the excitation survives. The second term, $\ensavg{c^\dag_k(0) c_k(t)}$, can be interpreted similarly but for holes instead of particles.

Information about the allowed energies for an excitation with momentum $k$ is contained in the spectral (Lehmann) representation of the Green's function, which we arrive at by Fourier transforming and expanding using a basis of eigenstates $\ket{n}$ with energies $\epsilon_n$ and partition function $Z$
\begin{eqnarray}
G^R(k, \omega) &=& \frac{1}{Z} \sum_{n,m} e^{-\beta \epsilon_n} \frac{\left| \matrixel{m}{c^\dag_k}{n} \right|^2}{\omega + \epsilon_n - \epsilon_m + i \eta} + e^{-\beta \epsilon_n} \frac{\left| \matrixel{m}{c_k}{n} \right|^2}{\omega - \epsilon_n + \epsilon_m + i \eta}\\
&=& G^+(k,\omega) + G^-(k,\omega). 
\end{eqnarray}
Here $\eta \in \mathbb{R}^+$ and we always consider the limit $\eta \to 0^+$. The terms $G^+$ and $G^-$ come from the anticommutator in $G^R$ and can again be associated with creating a particle and hole excitation respectively.

We define the single-particle spectral function as
\begin{eqnarray}
A(k, \omega) &=& -\frac{1}{\pi} \text{Im} \{G^R(k, \omega) \}\\
&=& \frac{1}{Z} \sum_{n,m} e^{-\beta \epsilon_n} \left| \matrixel{m}{c^\dag_k}{n} \right|^2 \delta \left(\omega + \epsilon_n - \epsilon_m \right) + e^{-\beta \epsilon_n} \left| \matrixel{m}{c_k}{n} \right|^2 \delta \left(\omega - \epsilon_n + \epsilon_m \right)\\
&=& A_\text{unocc}(k, \omega) + A_\text{occ}(k, \omega)\\
&=& A(k, \omega) \left[1 - F(\omega) \right] + A(k, \omega) F(\omega).
\end{eqnarray}
$A_\text{unocc}$ represents particle injection and $A_\text{occ}$, the occupied spectral function, represents photoemission. 

The spectral function counts single-particle excitations at a given momentum and frequency. In this sense it may be thought of as a momentum resolved density of states. More precisely, we have the following relationship between the density of states, $\rho(\omega)$ and the spectral function
\begin{eqnarray}
\rho(\omega) &=& \int dk \ A(k, \omega) \label{eq:dos}.
\end{eqnarray}

The spectral function is also closely related to the momentum distribution function
\begin{eqnarray}
\ensavg{n_k} &=& \ensavg{c^\dag_k c_k} = \int_{-\infty}^\infty d\omega \ A(k, \omega) F(\omega) \label{eq:momentum_dist}.
\end{eqnarray}

Finally, the spectral function is normalized to unity
\begin{eqnarray}
\int_{-\infty}^\infty d\omega \ A(k, \omega) &=& 1 \label{eq:spec_fn_norm}.
\end{eqnarray}

\section{Linear Response}

The Hamiltonian of our system is given by
\begin{eqnarray}
\mathcal{H} &=& \mathcal{H}_o + \mathcal{H}'\\
\mathcal{H}' &=& \Omega \sum_k \left(  c^\dag_{k\uparrow} c_{kf} + c^\dag_{kf} c_{k\uparrow} \right)\\
\mathcal{H}_o &=& \sum_{\sigma k} \epsilon_k n_{k\sigma} + U \sum_i n_{i\uparrow} n_{i\downarrow},
\end{eqnarray}
where $\mathcal{H}'$ describes the radiofrequency probe with Rabi frequency $\Omega$ and $\mathcal{H}_o$ is the Fermi-Hubbard Hamiltonian on a square lattice with $\epsilon_k = -2t \left[ \cos(k_x) + \cos(k_y) - 2 \right]$. We choose the band minimum of the non-interacting dispersion as zero energy. This implies that half-filling occurs at $\mu/t = 4 + U/2$. 

In the experiment, we measure the total density of atoms transferred by the perturbation portion of the Hamiltonian acting for some time $\delta t$ on our system. Assuming that the time is short enough and the perturbation is weak enough, we can apply linear response theory and the atom density is proportional to the atom current. The atom current operator is given by
\begin{eqnarray}
\dot{n}_{kf} &=& \frac{i}{\hbar} \left[\mathcal{H}', n_{kf} \right] = -\frac{i}{\hbar} \Omega \left(c^\dag_{kf} c_{k\uparrow} - c^\dag_{k\uparrow} c_{k f} \right).
\end{eqnarray}

The expectation value of the momentum resolved current for rf photon energy $h\nu$ is given by the Kubo formula as
\begin{eqnarray}
I(k, h\nu) &=& \ensavg{\dot{n}_{kf}(t)}_\nu = \Omega \int_{-\infty}^\infty dt' \ G^R(k, t - t')\\
G^R(k, t-t') &=& - \frac{i}{\hbar} \theta(t-t') \ensavg{ \left[ \dot{n}_{kf}(t), \mathcal{H}'(k, t')  \right]}.
\end{eqnarray}

Supposing that the final state $\ket{f}$ is unoccupied and does not interacting with the initial state $\ket{\uparrow}$, this expression can be simplified \cite{Chen2009, Toermae2016} to arrive at
\begin{eqnarray}
I(k, h\nu) &=& A(k, \epsilon_k - h\nu - \mu) F(\epsilon_k - h\nu - \mu) \label{eq:current},
\end{eqnarray}
where $\mu$ is the chemical potential of the interacting system, and $A$ is the spectral function of the interacting system described by $\mathcal{H}_o$. This expression assumes that the energy of the final state $\ket{f}$ is larger than the initial state. If the energy of the final state is smaller, then we should replace $h\nu$ with $-h \nu$ in the above expression.

In this expression $\epsilon_k$ is the energy of the outcoupled atom, implying that $\epsilon_k - h \nu$ is the initial energy. We must additionally subtract $\mu$ from the argument because the spectral function is defined such that $\omega = 0$ represents the system at the chemical potential.

This expression implies that the momentum resolved current is a measure of the occupied spectral function, however the presence of $\epsilon_k$ in the energy argument means that a measurement at a single frequency $\nu$ is a complicated surface in $k - \omega$ space.

We can rewrite eq.~\ref{eq:current} to give us the occupied spectral function at equal $\omega$,
\begin{eqnarray}
A(k, \omega) F(\omega) &=& I(k, \epsilon_k - \omega - \mu) \label{eq:spec_fn_from_current}.
\end{eqnarray}

\section{Trap averaging}

Due to the harmonic trapping potential, the density of our Hubbard system develops a spatial dependence. Working in the semiclassical local density approximation (LDA), we model this as an effective spatial variation of the chemical potential according to $\mu_i = \mu - V_i$, where $V_i$ is the potential at site $i$. The current is now given by
\begin{eqnarray}
I(k, h \nu) &=& \sum_i \ A(k, \epsilon_k - h \nu - \mu_i, \mu_i ) F(\epsilon_k - h \nu - \mu_i),
\end{eqnarray}

where we explicitly include the dependence of $A$ on $\mu$.

Determining the spectral function for the trapped system is more complicated than in the uniform system (eq.~\ref{eq:spec_fn_from_current}) because the convention is to reference $\omega$ to the chemical potential. This choice of reference is of little consequence when working at fixed density, but when working with multiple densities we must use a consistent reference point. In this case the physically meaningful quantity is $\omega + \mu_i$. In the trapped system our energy reference is the chemical potential (or the central chemical potential in LDA language), meaning that $\omega + \mu_i - \mu$ becomes the argument of our spectral functions
\begin{eqnarray}
I(k, \epsilon_k - \omega - \mu) &=& \sum_i A \left(k, \omega + \mu_i - \mu, \mu_i \right) F(\omega + \mu_i - \mu ). 
\end{eqnarray}

As we expect eq.~\ref{eq:current} to hold for the spectral function for the trap $A_\text{trap}$, we identify
\begin{eqnarray}
A_\text{occ, trap}(k, \omega) &=& \frac{ \sum_i A \left(k, \omega + \mu_i - \mu, \mu_i \right) F(\omega + \mu_i - \mu) }{ \sum_i 1}\\
A_\text{trap}(k, \omega) &=& \frac{ \sum_i A \left(k, \omega + \mu_i - \mu, \mu_i \right) }{ \sum_i 1},
\end{eqnarray}

where normalizing by the total number of sites in the trap ensures that the eqs.~\ref{eq:dos}, \ref{eq:momentum_dist}, and \ref{eq:spec_fn_norm} continue to hold.

In the experiment we have direct access to the density on each site, $n_i$. Therefore, using DQMC we effectively compute $\mu(n)$, the chemical potential as a function of the atomic density.

We compute the spectral function at the densities $n = 0.2 - 1.8$ in steps of $0.2$ and for each density in the trap we use the closest computed density to determine the chemical potential.

\section{EDCs experiment to theory comparison}

\begin{figure}
\includegraphics[width=\textwidth]{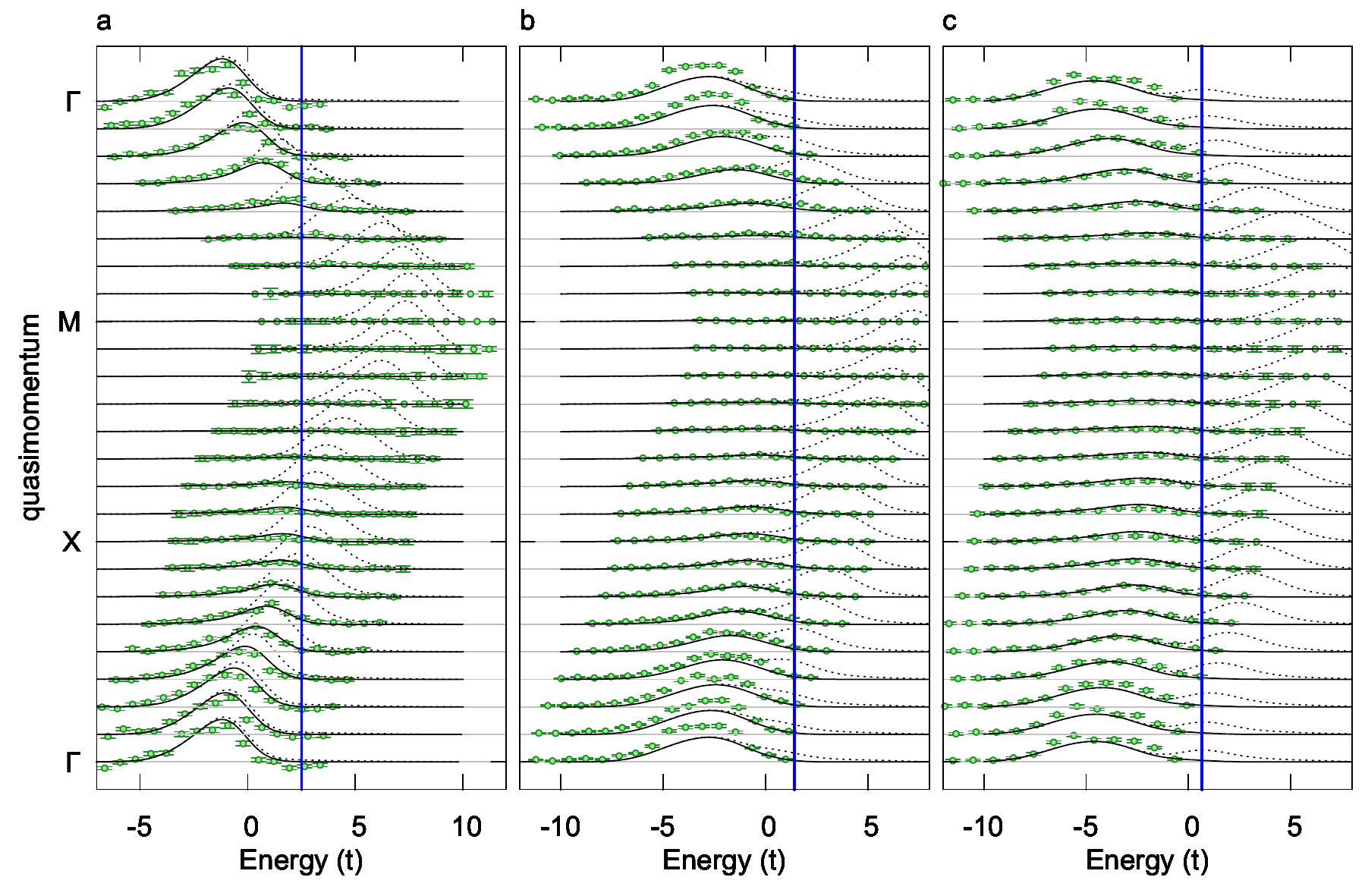}
\caption{ {\bfseries EDCs versus interaction.}  {\bfseries a}, EDCs from the experimental occupied spectral function (green points) and theory occupied spectral function (black solid lines) and theory spectral function (black dashed lines) for experiment at $U/t = -3.7(1)$ and $T/t = 0.48(2)$ and theory $U/t = -4$ and $T/t = 0.5$. {\bfseries b}, experiment $U/t = -6.0(1)$ and $T/t = 0.50(2)$ and theory $U/t = -6$ and $T/t = 0.5$. {\bfseries c} experiment $U/t = -7.5(1)$ and $T/t = 0.55(3)$ and theory $U/t = -7.5$ and $T/t = 0.55$.} 
\label{fig:edcs_vs_int}
\end{figure}

\begin{figure}
\includegraphics[width=\textwidth]{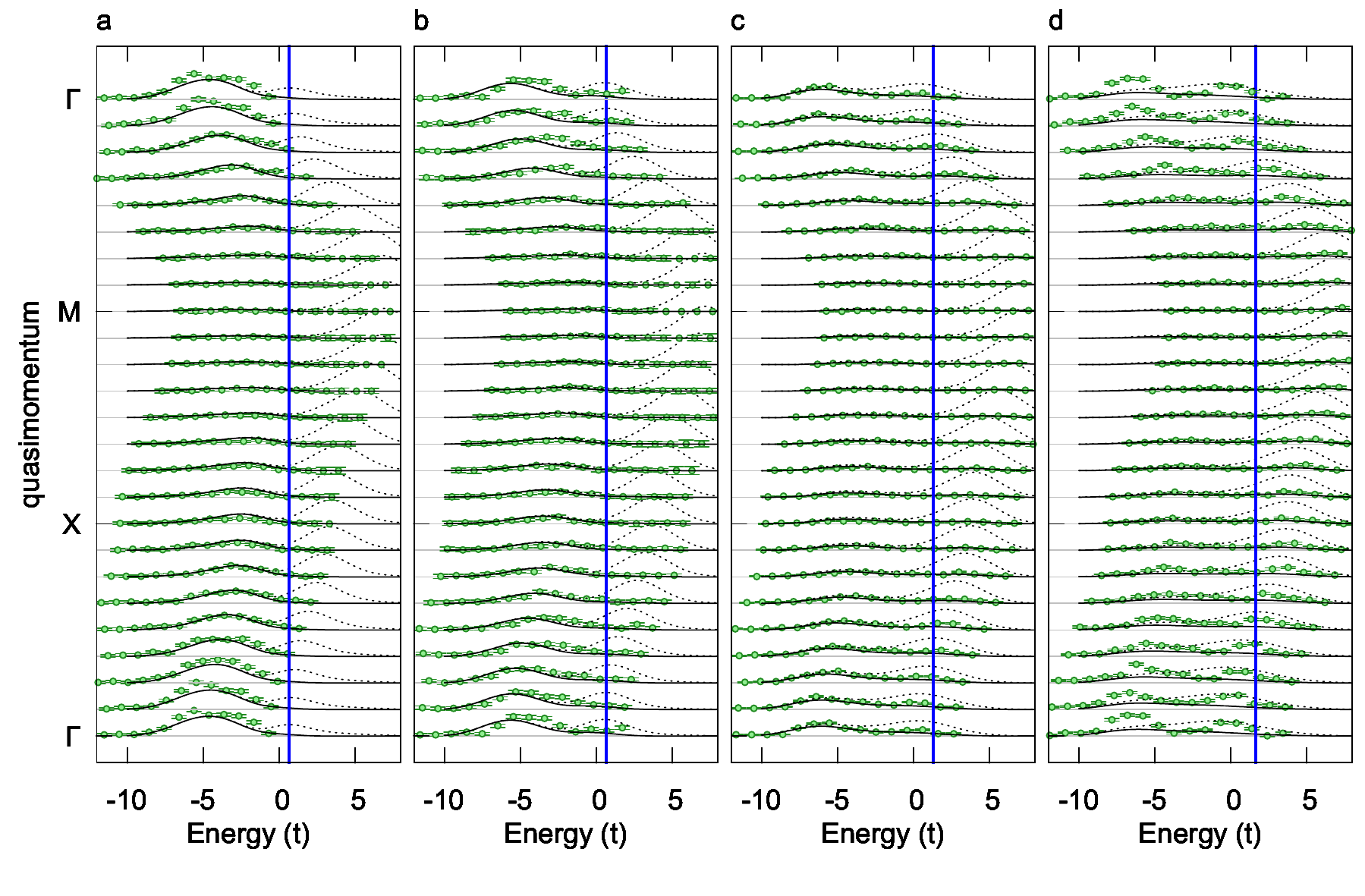}
\caption{ {\bfseries EDCs versus temperature.}  {\bfseries a}, EDCs from the experimental occupied spectral function (green points) and theory occupied spectral function (black solid lines) and theory spectral function (black dashed lines) for experiment at $U/t = -7.5(1)$ and $T/t = 0.55(3)$ and theory at $U/t = -7.5$, $T/t = 0.55$. {\bfseries b}, experiment $U/t = -7.5$ and $U/t = 0.93(5)$ and theory at $U/t = -8$ and $T/t = 1.04$. {\bfseries c}, experiment $U/t = -7.5$ and $T/t = 2.5(1)$ and theory $U/t = -8$ and $T/t = 2.83$. {\bfseries d}, experiment $U/t = -7.5$ and $T/t = 4.5(1)$ and theory $U/t = -8$ and $T/t = 5.0$.} 
\label{fig:edcs_vs_temp}
\end{figure}

It is instructive to compare the energy distribution curves for the experiment and theory data shown in the main text. Presenting the data in this way allows a direct comparison of theory and experiment. Here we present the EDCs corresponding to fig.~\ref{fig:vs_interaction} in the main text in fig.~\ref{fig:edcs_vs_int}, and the EDCs corresponding to fig.~\ref{fig:vs_temp} in the main text in fig.~\ref{fig:edcs_vs_temp}.

\FloatBarrier
 
 \FloatBarrier
 
\section{Full spectral functions}

Determining the presence of a pseudogap in the experimental system is complicated by the fact we measure the occupied spectral function. However, in the theory we can directly see the pseudogap open in the full spectral function. Here we present the full trap averaged spectral functions corresponding to the occupied spectral functions shown in figs.~\ref{fig:vs_interaction} and \ref{fig:vs_temp}. Here, unlike in the main text, we do not use any Gaussian broadening.

\begin{figure}
\includegraphics[width=\textwidth]{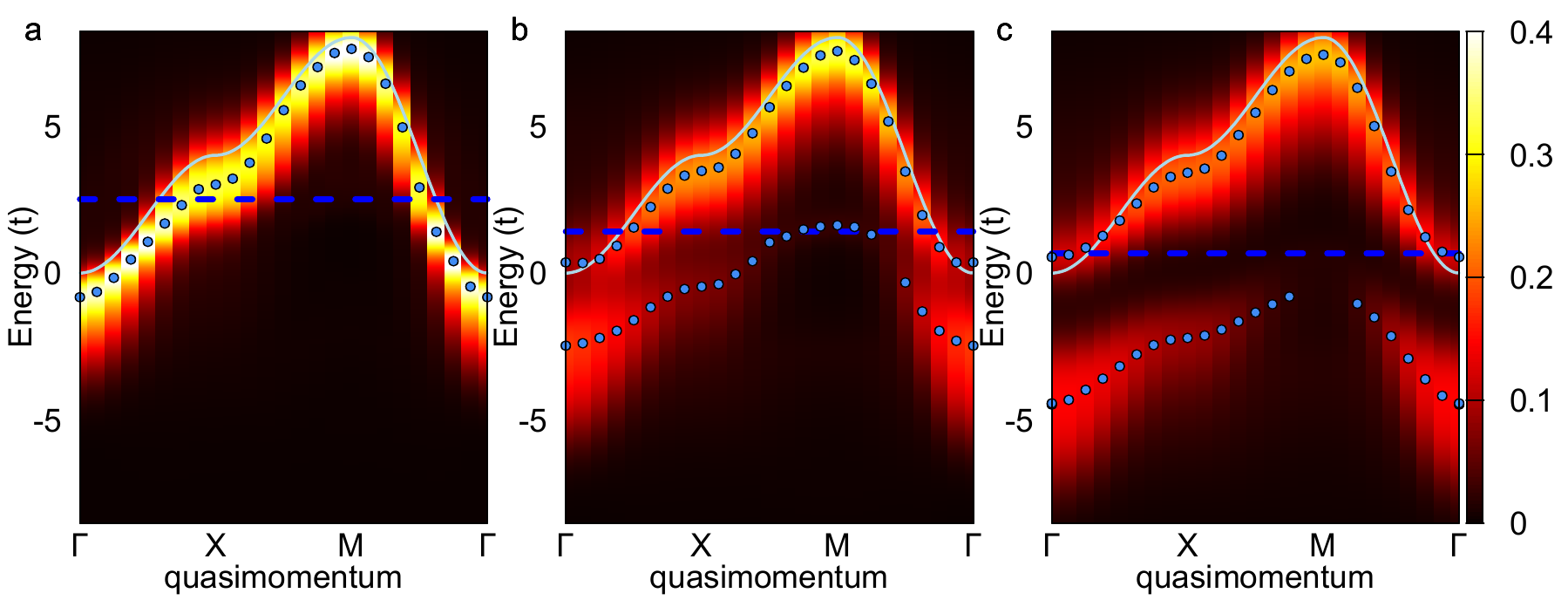}
\caption{ {\bfseries Spectral function versus interaction.}  {\bfseries a}, DQMC trap average spectral function for $U/t = -4$ and $T/t = 0.5$. {\bfseries b}, $U/t = -6$ and $T/t = 0.5$. {\bfseries c}, $U/t = -7.5$ and $T/t = 0.55$.} 
\label{fig:full_specfn_vs_interaction}
\end{figure}

In fig.~\ref{fig:full_specfn_vs_interaction}, a pseudogap is visible for $U/t = -6$ and $-8$ at lower temperatures $T/t \sim 0.5$. No pseudogap is visible for $U/t = -4$. This figure corresponds to fig.~\ref{fig:vs_interaction} in the main text.

\begin{figure}
\includegraphics[width=\textwidth]{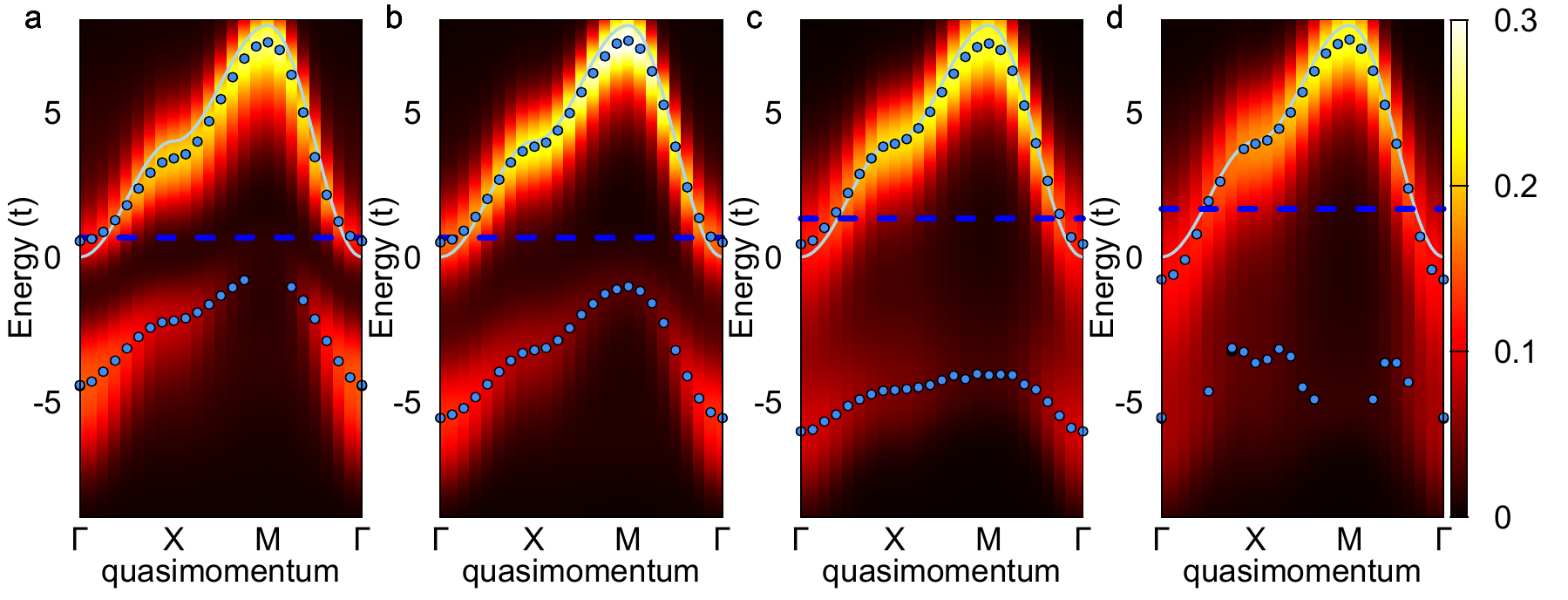}
\caption{ {\bfseries Spectral function versus temperature.}  {\bfseries a}, DQMC trap average spectral function for $U/t = -7.5$ and $T/t = 0.55$. {\bfseries b}, $U/t = -8$ and $T/t = 1.04$. {\bfseries c}, $U/t = -8$ and $T/t = 2.83$. {\bfseries d}, $U/t = -8$ and $T/t = 5.0$.} 
\label{fig:full_specfn_vs_temp}
\end{figure}

In fig.~\ref{fig:full_specfn_vs_temp} a pseudogap is visible for $T/t < 3$ and appears to have disappeared by $T/t = 5$. This figure corresponds to fig.~\ref{fig:vs_temp} in the main text.

\FloatBarrier

\FloatBarrier

\section{Dispersion from $A(k,\omega)$ versus $A_\text{occ}(k, \omega)$}

\FloatBarrier

\begin{figure}
\includegraphics[width=7cm]{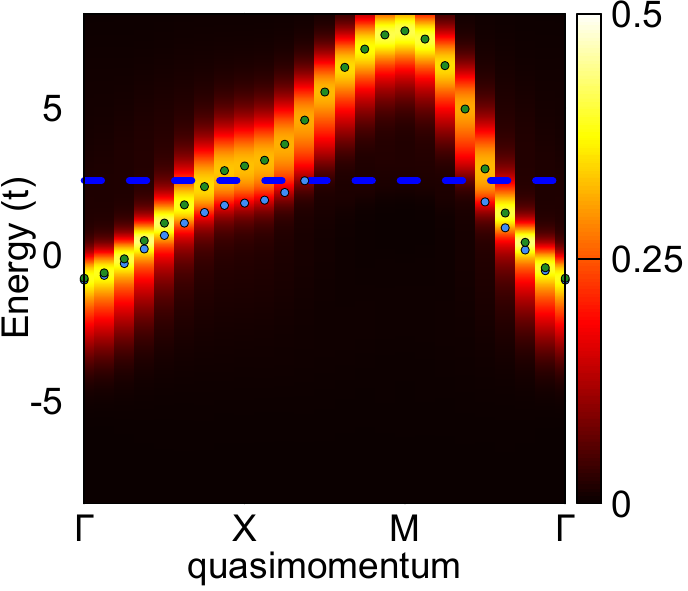}
\caption{ {\bfseries Dispersion from $A(k, \omega)$ versus $A_\text{occ}(k, \omega)$.} The full spectral function along high symmetry points in the Brillouin zone and the chemical potential (blue line), the points of peak spectral weight for $A(k, \omega)$ (green points) and $A_\text{occ}(k, \omega)$ (blue points).} 
\label{fig:dispersion_comparison}
\end{figure}

The dispersion inferred by determining the peak in spectral weight at each momentum is significantly affected by the presence of the Fermi function. Near the chemical potential this causes the dispersion of the occupied spectral function to differ significantly from the real dispersion. This effect is most pronounced outside of the pseudogap regime because in that case there is significant spectral weight at the chemical potential. To illustrate this effect, we perform the peak-finding procedure described in the text using both the full and occupied spectral functions. We show only peaks with heights greater than $0.02$. The results are shown in fig.~\ref{fig:dispersion_comparison}.

\FloatBarrier

\section{DQMC calculation of the single-particle spectral function}
We perform DQMC simulations on the attractive Hubbard model \cite{BSS,White1989} with interaction strengths and temperatures $U/t=-3.7, T/t=0.48$; $U/t=-6, T/t=0.5$; $U/t=-7.5; T/t=0.55$; $U/t=-8, T/t=0.44$; $U/t=-8, T/t=1.04$; $U/t=-8, T/t=1.2$; $U/t=-8, T/t=2.83$; and $U/t=-8, T/t=5$. For each parameter set, we consider filling levels of $\ensavg{n} = 0.2, 0.4, 0.8,$ and $1$. The chemical potential is tuned to achieve the desired doping level to within an accuracy of $O(10^{-4})$. The imaginary time interval $[0,\beta]$ is discretized into steps of around $0.05$ or less, resulting in negligible Trotter errors for our simulations. All simulations were on square clusters of size $16 \times 16$. We run $160$ independently seeded Markov chains with $1000$ warmup and $5000$ measurement sweeps each, giving a total of $8 \times 10^5$ sweeps for each doping and temperature. Unequal time measurements are performed on every other sweep.

To ensure numerical stability in computing the equal-time Green's functions, we use the prepivoting stratification algorithm as described in \cite{Tomas2012}, allowing up to 10 matrix multiplications before performing a QR decomposition. The unequal time Green's functions are constructed using the Fast Selected Inversion algorithm described in \cite{Jiang2016}, with blocks corresponding to the product of matrices from 10 time steps.

We perform maximum entropy analytic continuation (MaxEnt) \cite{Jarrell1996} to extract the single particle spectral function from the imaginary time Green's functions measured in DQMC. The MaxEnt optimization utilizes Bryan's algorithm as described in \cite{Jarrell1996}. For the choice of model function, we use a non-informative flat function, and select the parameter $\alpha$ using the prescription given in \cite{Bergeron2016}. We have checked that the results are nearly identical when using the Classic formulation of MaxEnt with an annealing method for model functions \cite{Jarrell1996}.

To determine the spectral function for densities above half-filling, we take advantage of the symmetry
\begin{eqnarray}
A(k, \omega, n) &=& A(k + Q, -\omega, 2-n),
\end{eqnarray}
where $Q = (\pi, \pi)$.

\section{Fitting the temperature and interaction}

\begin{figure}
 \includegraphics[width=\textwidth]{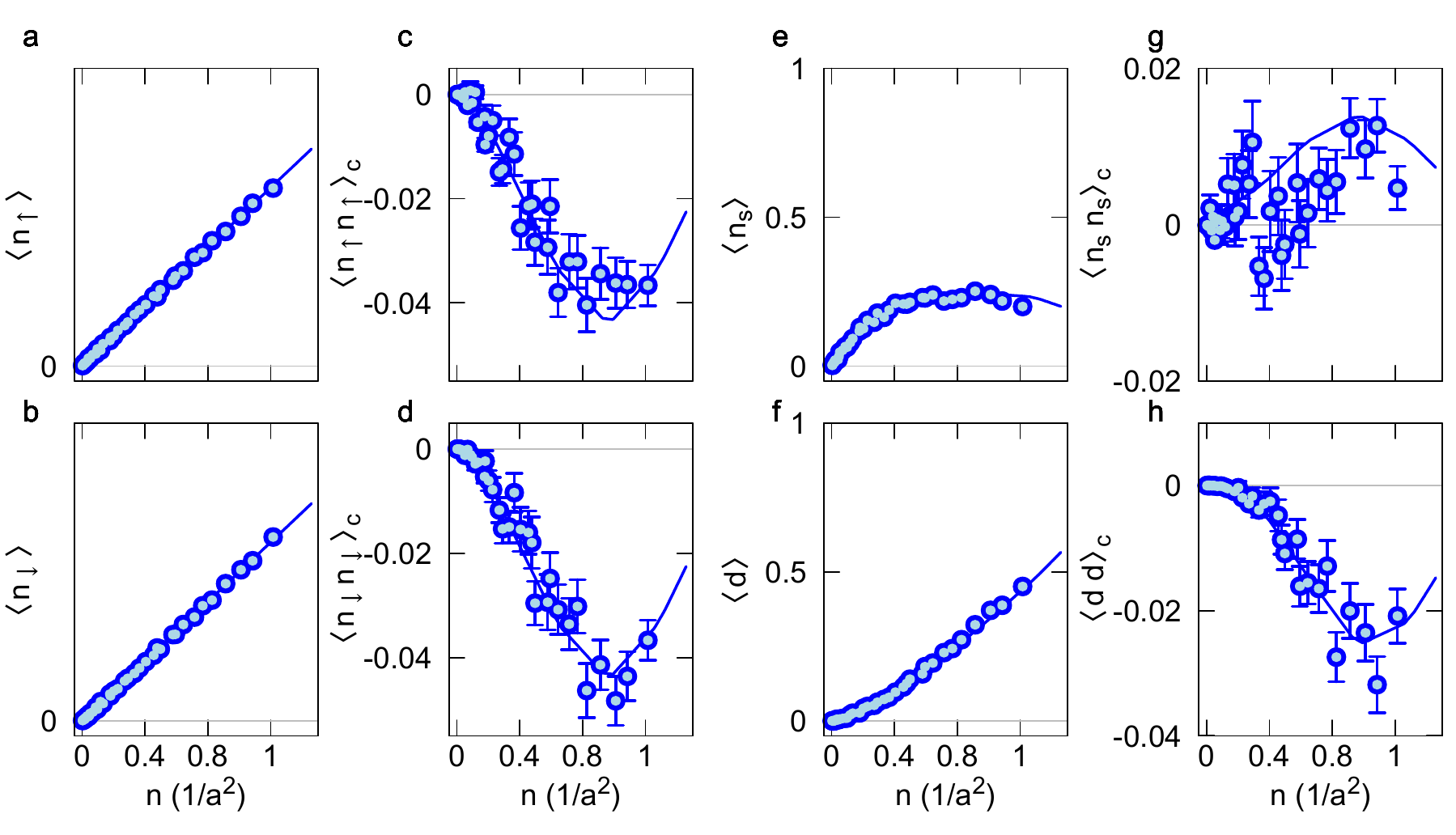}
 \caption{ {\bfseries DQMC parameter fitting.}  Experimental data (points) and DQMC fit (line) for various quantities versus total density. {\bfseries a}, spin up density. {\bfseries b}, spin down density. {\bfseries c}, nearest-neighbor spin up correlator. {\bfseries d}, nearest-neighbor spin down correlator. {\bfseries e}, singles density. {\bfseries f}, doubles density. {\bfseries g}, nearest-neighbor singles correlators. {\bfseries h}, nearest-neighbor doubles correlator.} 
 \label{fig:temp_fit}
\end{figure}

To determine the interaction, temperature, and chemical potential of our gas we perform a simultaneous fit of the quantities $C_\uparrow(1)$, $C_\downarrow(1)$, $n_s$, $C_s(1)$, $d$, $C_d(1)$ versus total density to DQMC data with $U/t$ and $T/t$ as free parameters. Here $C_\uparrow(1) = \ensavg{n_\uparrow(i_x, i_y) n_\uparrow(i_x + 1, i_y)} - \ensavg{n_\uparrow(i_x, i_y)} \ensavg{n_\uparrow(i_x + 1, i_y)}$ and similarly for the other correlators. By fitting these various correlators versus density we remove the need to precisely character the trapping potential by taking advantage of the known equation of state. The procedure is similar to that described in \cite{Brown2019}, but using additional quantities such as the doublon-doublon correlator and the single-single correlator allows simultaneous determination of both $U/t$ and $T/t$ and leads to higher precision fitting. The result of an exemplary fit is shown in fig.~\ref{fig:temp_fit}. In this figure, the DQMC data has been corrected using imaging fidelities and the experimental correlators are the values measured in the experiment. We used a doublon transfer efficiency of $0.9$ and an imaging fidelity of $0.96$. Note that the doublon imaging efficiency is the product of these two efficiencies. Densities are corrected by a single factor of the relevant efficiency, and correlators are corrected by the efficiency squared. The fit in fig.~\ref{fig:temp_fit} yields a temperature of $T/t = 0.38$ and interaction of $U/t = -4.2$.

\section{Band mapping simulation}

As our band mapping time, \SI{200}{\micro \second}, is not small compared with the quarter period expansion time of \SI{700}{\micro \second}, we perform a 1D simulation of the procedure following \cite{McKay2009} to check what effect atomic motion during the band mapping has on the final density profiles. 

We first find the lowest eigenstates of a single-particle Hamiltonian with potential given by the sum of a lattice and harmonic trap in real space. Then we change the lattice depth linearly from the initial depth to zero while time evolving the initial eigenstates using the Crank-Nicholson method. We simulate the quarter period expansion in the harmonic trap evolution using the propagator

\begin{equation}
U(x, t | x', 0) = \sqrt{ \frac{1}{2 \pi i \sin(\omega t) l^2} } \exp \left[i \frac{1}{2l^2 \sin(\omega t) } \left( (x^2 + x'^2) \cos(\omega t) - 2 x x'\right) \right],
\end{equation}
where $l = \sqrt{\frac{\hbar}{m\omega}}$ is the harmonic oscillator length.

\begin{figure}
 \includegraphics[width=0.5\textwidth]{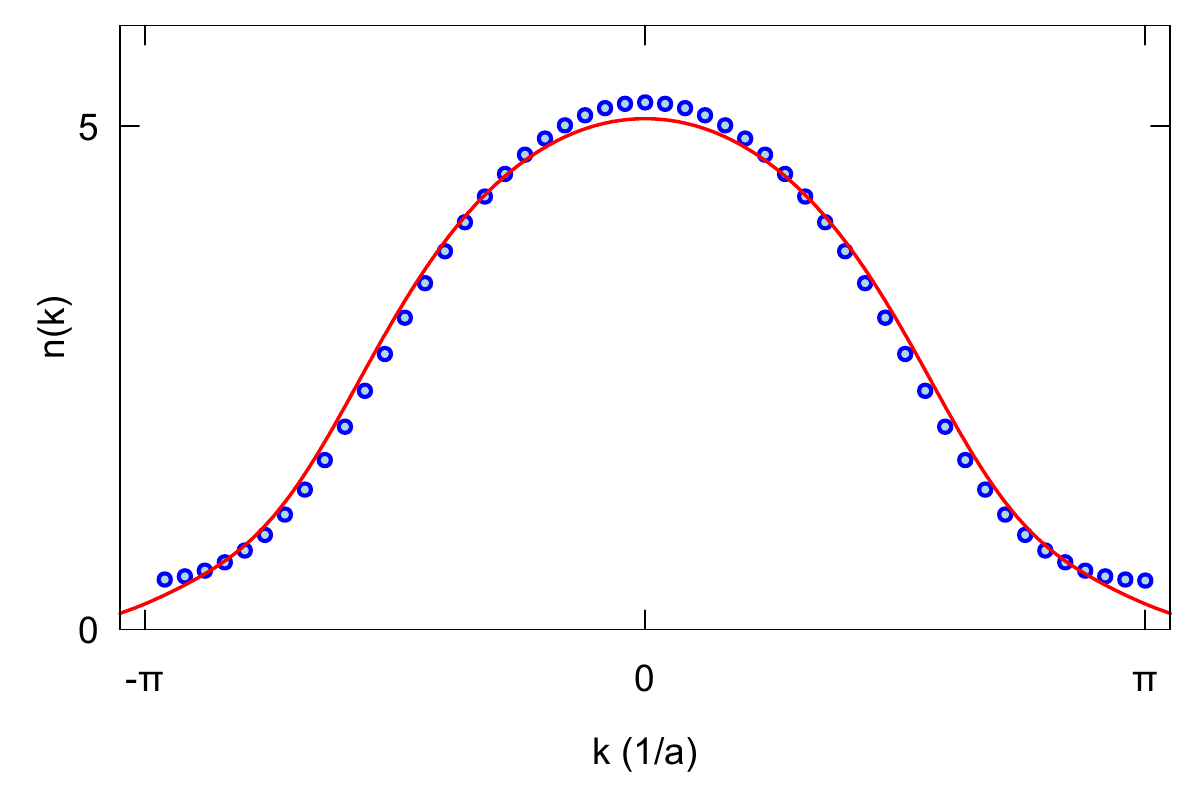}
 \caption{ {\bfseries Band mapping simulation.} Quasimomentum distribution (blue circles) and scaled atomic density after band mapping procedure (red line).} 
 \label{fig:band_map}
\end{figure}

The results of this procedure are shown in fig.~\ref{fig:band_map} for an initial temperature of $T/\si{\erecoil} = 1/30$, a lattice depth of \SI{6.5}{\erecoil}, a harmonic trapping frequency of $\omega_i = (2\pi) \SI{200}{\hertz}$ and a band mapping time of $\SI{200}{\micro \second}$. The expansion trap frequency is $\omega_f = (2\pi) \SI{360}{\hertz}$ and one quarter period is $ \sim \SI{700}{\micro \second}$.

We compare the results of the band mapping procedure to the initial quasimomentum distribution, which we determine by projecting the eigenstates of the initial trap on to the Bloch eigenstates of the lattice. We determine the scaling factor between the two using a least-squares fit as we do for the experimental data. This fit typically requires less than a $10 \%$ rescaling of the density.  We find that the largest discrepancy occurs at the edges of the Brillouin zone where the gaps are smallest. The average fractional discrepancy between the quasimomentum distribution and the scaled band map density distribution is less than $10 \%$. 

In the experiment the trap  parameters are $\omega = (2\pi) \SI{160}{\hertz}$, $t/h = \SI{1400}{\hertz}$, and \SI{5}{\erecoil} at $U/t = -4$; $\omega = (2\pi) \SI{185}{\hertz}$, $t/h = \SI{1230}{\hertz}$, and \SI{5.5}{\erecoil} at $U/t = -6$; and $\omega = (2\pi) \SI{200}{\hertz}$, $t/h = \SI{1000}{\hertz}$, and \SI{6.5}{\erecoil} at $U/t = -8$.

\section{Magnetic field stabilization}

To reach \SI{120}{\milli \gauss} we null Earth's field in the $x$ and $y$ directions using stable current sources. Due to the high magnetic field sensitivity of the $\ket{2} - \ket{3}$ rf transition at low fields, we must actively stabilize the field in the $z$ direction. We measure the magnetic field using a Stefan Mayer FLC100 fluxgate sensor \cite{Dedman2007}, and control it using a Helmholtz coil of radius $\sim \SI{0.5}{\meter}$ and a home built servo driving a power MOSFET.

We test the stability of our field by measuring the width of the $\ket{2} - \ket{3}$ transition using a polarized gas of $\ket{3}$ atoms in a deep lattice, ($\sim \SI{60}{\erecoil}$), which freezes out tunneling. We fit the spectra to a Lorentzian line shape and find a half width half max of $\sim \SI{300}{\hertz}$. The full width corresponds to a field fluctuation of $\SI{0.3}{\milli \gauss}$.

We assess the long term stability of our field by measuring the resonance frequency of the polarized gas before and after each experimental data set. We see frequency drifts of $\sim \SI{1}{\kilo \hertz}$ (\SI{0.5}{\milli \gauss}), which is of the same order as our Fourier broadening.

\end{document}